%% file: cair2018workshop-paper.tex
\documentclass[sigconf]{acmart}
\usepackage{booktabs} 
\usepackage{tikz,lipsum,lmodern}
\usepackage[most]{tcolorbox}
\usepackage{multirow}
\usepackage{multicol}
\usepackage{rotating}
\setcopyright{rightsretained}

\acmDOI{}

\copyrightyear{2018} 
\acmYear{2018} 

\acmConference[CAIR'18]{2nd International ACM SIGIR Workshop Conference on Conversational Approaches to IR}{July 13, 2018}{Ann Arbor, MI, USA}
\acmBooktitle{2nd International ACM SIGIR Workshop Conference on Conversational Approaches to IR, July 13, 2018, Ann Arbor, MI, USA}
\acmISBN{}

\begin{document}
\title{A Conceptual Framework for \\ Conversational Search and Recommendation}
\subtitle{Conceptualizing Agent-Human Interactions During the Conversational Search Process}

\author{Leif Azzopardi}
\affiliation{%
  \institution{University of Strathclyde}
  \city{Glasgow}
  \state{Scotland, UK}
}
\email{leifos@acm.org}
\author{Mateusz Dubiel}

\affiliation{%
  \institution{University of Strathclyde}
  \city{Glasgow}
  \state{Scotland, UK}
}
\email{mateusz.dubiel@strath.ac.uk}

\author{Martin Halvey}
\affiliation{%
  \institution{University of Strathclyde}
  \city{Glasgow}
  \state{Scotland, UK}
}
\email{martin.halvey@strath.ac.uk}

\author{Jeffrey Dalton}
\affiliation{%
  \institution{University of Glasgow}
  \city{Glasgow}
  \state{Scotland, UK}
}
\email{jeff.dalton@glasgow.ac.uk}

\begin{abstract}
The conversational search task aims to enable a user to resolve information needs via natural language dialogue with an agent. In this paper, we aim to develop a conceptual framework of the actions and intents of users and agents explaining how these actions enable the user to explore the search space and resolve their information need. We outline the different actions and intents, before discussing key decision points in the conversation where the agent needs to decide how to steer the conversational search process to a successful and/or satisfactory conclusion. Essentially, this paper provides a conceptualization of the conversational search process between an agent and user, which provides a framework and a starting point for research, development and evaluation of conversational search agents.
\end{abstract}


\maketitle
\input{introduction}
\input{related-work}

\input{simulated-conversation}

\input{user-actions}
\input{agent-actions}
\input{decision-points}
\input{closing}

\bibliographystyle{ACM-Reference-Format}
\bibliography{cairbib}
\end{document}

%% file: introduction.tex
\section{Introduction}\label{sec_introduction}
A conversational search agent needs to support the user in finding, exploring, and understanding the 
possible options and information objects that are available --- which will help to satisfy the user's information need. While past work has started to tease out different actions that users and agents perform and respond to during the conversational search process, there has been little work on formalizing these actions and decisions. Thus the goal of this paper is to develop a conceptual framework of different actions and intents, along with the key decision points within the conversation. Our aim is to make these tasks explicit in order to formalize the research, development and evaluation of conversational search agents. To this end, we first examine 
the key actions and intents identified in past work, and enumerate these along with others that can be naturally inferred from a simulated conversational context, before discussing the key decisions that the agent needs to make in order to advance the conversation to a satisfactory or successful end. 

%% file: related-work.tex
\subsection{Background and Related Work}\label{sec_related_work}
To ground the proposed framework we first provide a brief overview of \textit{conversational agents} in general. Then, we focus in on \textit{conversational search agents} and describe what actions various proponents suggest that they need to be able to perform and accomplish. 

Liu et al. \cite{zhanyiliuzhengyuniujianyunniehuawuhaifengwang2017} provide a high-level overview of six abilities a conversational agent needs to possess: 
\begin{enumerate}
\item  \textit{filtering out superfluous information} e.g. fillers, pauses, false starts; 
\item  \textit{determining an appropriate system response} and the need for sophisticated decision making mechanism (e.g. should the agent show results, show a hierarchy, ask a follow up query?, etc.); 
\item  \textit{answer aggregation} to present a summary answer rather than a just ranked list of results;
\item  \textit{conversation management} that considers and maintains the search goals, conversational history and current state of the agent's understanding;
\item  \textit{general knowledge} that the agent should have about external world in order to efficiently exploit contextual information and correctly process the user's query; 
\item  and \textit{personality and moral responsibility} to respond to emotional and sensitive queries. 
\end{enumerate}
Similarly, Allen et al. ~\cite{allen2001architecture} suggest that a conversational agent needs to be aware of the current state of the conversation, and continuously update its representation given the user's responses in order to generate the best response. A key action that an agent needs to perform is elicitation~\cite{christakopoulou2016towards}, where the agent needs to learn about the user's preferences. Then the agent needs to be able to help the user to efficiently explore the search space ~\cite{winterboer2011user}. Demberg et al. ~\cite{demberg2011strategy} suggests that agents also need to be able to cope with both under and over-constrained requests.

In developing a conversational agent, Moore~\cite{robertj.moore2018} suggests that an agent needs to have strategies for repair and disengagement. For example, the agent needs to be able to repeat information or ask for information to be repeated in case of misunderstanding, as well as recognize when the conversation is closing, correcting the recipient and dealing with insults. Bohus and Rudnicky \cite{bohus2005sorry} also point out that such agents need to be able to handle misunderstandings (and be able to ask for a statement to be rephrased etc., but also that agents needs to be able to explain their current state (i.e. report what they understand)) - so that misunderstandings can be corrected. Moore~\cite{robertj.moore2018} postulates that it is crucial for an agent to be able to perform such actions to exhibit basic conversational competence. 

Various taxonomies of conversational actions have been developed in the context of dialogue management. For instance, Henderson et al.~\cite{henderson2013dialog} in their `Dialogue State Tracking Challenge' list: \textit{affirm}, \textit{confirm-domain}, \textit{negate} and \textit{repeat} among the examples of agent actions and \textit{deny}, \textit{inform} and \textit{request} as some of the users actions.  In Clark and Schaefer~\cite{clark1989contributing}, they present several methods for establishing common ground in conversation, these include: \textit{continued attention}, \textit{next contribution}, \textit{acknowledgement}, \textit{demonstration} and \textit{display}. These have also been formalized as part of an ISO standard for semantic annotation of dialogues, ISO 24617~\cite{Bunt2010TowardsAI}, more commonly referred to as dialogue act markup language (DiAML).  While the above actions and taxonomies can be applied to any general dialogue, in this paper, we focus on actions and intents that are relevant to the conversational search process. 

More specifically related to conversational search, Radlinski and Craswell \cite{radlinski2017theoretical} provide  a theoretical framework, that puts forward five properties that a search agent needs to have in order to be conversational. The properties are: 
\begin{enumerate}
\item \textit{user revealment} where the user discloses to the agent their information needs, 
\item \textit{agent revealment} where the agent reveals what the agent understands, what actions it can perform, and what options are available to the user,
\item \textit{mixed initiative} where both the agent and the user can initiative and direct the conversation, 
\item \textit{memory} where the agent tracks and manages the state of the conversation, the user's information need, etc., and, 
\item \textit{set retrieval} where agent needs to be able to work with, manipulate and explain the sets of options/objects which are retrieved given the conversational context.
\end{enumerate} 

The above properties are required so that the agent can facilitate search by helping user to formulate their information need and build expectations regarding its capabilities. During the search process, the agent takes initiative and uses memory to retain information relevant to the query. Before presenting results back to user, the agent needs to reason about the utility of retrieved information and decide on what to present to the user. 

Trippas et al. \cite{trippas2018informing} provide a different high-level formalization of the conversational search process distinguishing three phases: (1) query formulation, (2) search results exploration and (3) query re-formulation. They note that during each phase, the agent needs to elicit details of the information need and obtain feedback on the results, and then use this to inform subsequent actions. They suggest that key activities that an agent should be able to perform include: listing and summarizing results and result pages, and requesting feedback from the user. 

Recently, we have seen a number of studies exploring how people interact with conversational search agents (where the agent is either a human with a search engine, or a simulated ``Wizard of Oz'' agent, e.g. \cite{vtyurina2017exploring,vtyurina2018exploring,wolters2009reducing,trippas2017people}.
In these studies they focus on ascertaining how people behave and interact with the agent, without imposing a defined interaction policy regarding how the ``agent'' should act~\cite{mcduff2017multimodal}. Instead, participants are provided with search tasks and roles but not explicitly instructed on how to complete them. Therefore, while they do not explicitly focus on specifying the actions taken by the agent (e.g. wizard or mediator) the conversational search logs could be useful in identifying common actions - and could be used to provide evidence to support the proposed framework.

%% file: simulated-conversation.tex
\section{Conversational Actions and Agent Decisions}
The main goal of this paper is to abstract out key actions and decisions points that manifest during the conversational search process. In order to do this, we will walk-through possible ways in which an agent and user may converse during the search process and draw out particular tasks that the agent needs to perform either to respond to the user or engage with the user. 

\textbf{Assumptions} We take a focused view of the conversational search setting, where we assume that our hypothesized agent wants to: 
\begin{enumerate}
\item {\bf help the user}: the agent is cooperative, seeks to serve the user's interests, is not adversarial and does not seek to maximize its own benefit, 
\item {\bf minimize the conversational effort}: the agent does not seek to waste time with idle chit-chat, 
and seeks to avoid burdening the user's cognitive capacity by asking excessive questions, or presenting irrelevant options, etc.,
and,
\item {\bf maximize the range/number of relevant options provided to the user}: the agent wants to provide the user with an understanding of the search space so the user can make an informed choice, as well as guide the user via continued exploration the search space subject to the time the user has available and the user's cognitive limits.
\end{enumerate}
Ultimately, we assume that the agent wants to: (i) help the user to resolve their information need, and/or (ii) help the user to understand the space of options available to the user. Depending on the domain of the conversational search agent, it could enable the search over very specific resources (e.g. flight, flowers, etc.) or be more general and enable ad-hoc and exploratory search of any kind of information objects (e.g. news, holidays, reviews, places, recipes, etc.).

Now, imagine that we have a user, with a goal in mind - maybe the goal is well-defined, precise and fixed (e.g. a known destination, time/date, etc.), or maybe it is loosely defined, vague and flexible (e.g. planning a holiday). This is unknown to the agent - and so the agent will need to discover what the user wants. The user will also have particular preferences. But again, we assume that these are, initially, unknown to the agent and will be revealed through the conversation(s).

While we will provide specific examples using natural language, it is also worth noting that we assume that the agent has sufficient capabilities to understand and respond to the requests of the user. Note that for the purposes of the simulated conversations we assume that the agent has the capacity to perfectly understand the user's utterances and requests, and can provide meaningful and relevant responses. We acknowledge that these are both very much open challenges that have yet to be fully addressed. Of concern here, is the second of these challenges i.e. a ``meaningful response'' which requires the agent needs to decide on what action(s) to take at any particular point in the conversational search process.

\textbf{Examples} Below we have two examples of conversations between users and agents, where the task is ad-hoc and exploratory in nature. The first considers the context of exploring holiday options, while the second is in the context of current affairs. These contexts present tasks that are open ended, ongoing and that evolve over the conversation and time.


    	\begin{center}
	\begin{tcolorbox}[colback=green!5!white,colframe=green!75!black,width=8cm]
    	\scriptsize
		\textbf{User}: I would like to go on holiday.\\
        \textbf{Agent}: Great, did you have a particular place in mind?\\
        \textbf{User}: I'd like a place that is sunny but also a lot of culture.\\
        \textbf{Agent}: Ok, how about visiting southern Europe, there are lots of cultural attractions there.\\
        ...\\
        \textbf{User}: That sounds good, what do you suggest?\\
        \textbf{Agent}: There are lots of place to pick from, so let's try and narrow it down. Would you prefer a city break, a beach holiday, a country vacation, or something else?\\
        ...\\
        \textbf{Agent}: If you are thinking about going to Tuscany in Italy, then you might also want to consider Galicia in Spain.
    
 \tcblower   
  
    \scriptsize
		\textbf{User}: I'm interested to know about countries that have taken efforts to help protect wildlife.\\
        \textbf{Agent}: Are you interested in general, or more recently? \\
        \textbf{User}: Umm, the recent efforts, please.\\
        \textbf{Agent}: Ok, most recently a wildlife sanctuary opened in the northern province of China, but a number of other countries such as Malaysia, Pakistan, Vietnam and several other countries have also been trying to protect wildlife.\\
        \textbf{User}: What efforts have Malaysia taken?\\
        \textbf{Agent}: They have recently developed a breeding program for the Sumatran Rhino, ... Malaysia has also tried to help other species which are endangered such as the Malayan Tiger and a bird called a Merak Pongsu.\\     
        ...\\
        \textbf{Agent}: You might also like to know about efforts in Thailand to help protect the Malayan Tiger.
    \end{tcolorbox}
    \end{center}  
    
During each of the above conversations, while in different contexts, we can start to see actions that are common to both, which can be abstracted out more generally. For example, in both cases, (i) a user \textit{discloses} part of their information need, (ii) the agent \textit{reveals} some details about the options available (given the partial information need), the agent \textit{elicits} more details regarding the user's information need, (iii) the user \textit{refines} or \textit{expands} their information need, (iv) the agent \textit{hypothesizes} about alternative information needs and then \textit{suggests} other options, and so forth. Of course, a user (and an agent) will communicate their intent to the other party through natural language (e.g. an utterance or chat). Another more general challenge is to identify from the natural language the intent and action that the user wants to communicate, and conversely how to express the intent and action of the agent to the user. 

In the next two subsections, we provide a \textit{non-exhaustive} list of the different kinds of actions users and agents might perform during the conversational search process (as shown in Table~\ref{tbl_actions}), before considering the different decisions agents will need to make during in the conversation. This list should be seen as a starting point
for conceptualizing agent-human interactions for conversational search. In creating this list we have attempted to represent the main conversational search actions previously observed and discussed in the literature, as well as other actions that can be naturally inferred from the conversational context. For simplicity we will provide examples based around planning a  holiday, though the same kinds of actions generalize to other search scenarios.

\textbf{Information Need Pathways} Before listing the different actions, we will assume that the conversational search process revolves around the agent trying to help the user resolve their information need. As such the agent will need to maintain a representation of the user's information need as it evolves over time. 

Given a point in the conversation, we refer to this representation as the \textit{Current Information Need} (CIN). It encapsulates what the agent has understood (and has modelled) given the preceding conversation. Given the CIN, the agent can retrieve the list of related information objects (i.e. the list of destinations, reviews, tours, etc that are returned by the query formulated based on the CIN). The agent will then use this list to inform and direct the subsequent actions. 

Over the course of the conversation, the CIN will change and evolve, and so the agent will also need to keep track of these {\it Past Information Needs} (PINs) and the corresponding set of associated objects. This is because the user may wish to refer back to a previous point in the conversation, or the agent may need to explain how they came to a particular point in the conversation. 

In addition to the user created information needs, the agent itself, may generate {\it Alternative Information Needs} (AINs) based on CIN and PINs. AINs are generated to provide other recommendations to the user that the user may not have explicitly considered, or not yet asked about. Again, these AINs have a corresponding space of associated information objects. The agent will also need to record what parts of the search space of objects has been revealed to the user over the course of the conversation.
\vspace{-6mm}



\begin{table}[]
\centering
\caption{An Overview of the Actions and Interactions.}
\vspace{-2mm}
\label{tbl_actions}
\begin{tabular}{|l|l|l|ll|l|l|}
\hline
                        \multicolumn{2}{|l}{} & \textbf{User} & \textbf{Agent} &                      \multicolumn{3}{l|}{}                \\
                                 \hline
& \multirow{10}{*}{ \begin{turn}{90}Query Formulation~\cite{trippas2018informing} \end{turn}} & \textbf{Reveal}      & \textbf{Inquire}      &  & \multirow{10}{*}{\begin{turn}{270}User Revealment~\cite{radlinski2017theoretical}\end{turn}} & \multirow{32}{*}{\begin{turn}{270}Memory~\cite{radlinski2017theoretical}\end{turn}}\\
&                                 & Disclose     &     &  &                     &               \\
&                                 & Non-Disclose     &     &  &                     &               \\
&									& Revise     &       &  &  &\\   
&									& Refine     &       &  &  &\\   
&                                 & Expand     &       &  &  &\\
&                                 &      & Extract       &  & & \\
&                                 &      &  Elicit     &  & & \\
&                                 &      &  Clarify     &  &  &\\
&                                  &      		&       &  & & \\
                               \cline{1-6}
\multirow{14}{*}{ \begin{turn}{90}Set Retrieval~\cite{radlinski2017theoretical}\end{turn}}& \multirow{14}{*}{ \begin{turn}{90}Result Exploration~\cite{trippas2018informing}\end{turn}} & \textbf{Inquire}      & \textbf{Reveal}      &  & \multirow{21}{*}{\begin{turn}{270}System Revealment~\cite{radlinski2017theoretical}\end{turn}} &\\
&                                 & List     		& List    &  &                              &      \\
&                                 & Summarize     	& Summarize      &  &  &\\   
&                                 & Compare     		& Compare      &  & & \\
&                                 & Subset     		& Subset      &  & & \\
&                                 & Similar     	& Similar       &  & & \\
                                 \cline{3-5}
&                                 & \textbf{Navigate}    		& \textbf{Traverse}      &  & &\\
&                                 &  Repeat    		& Repeat      &  &  &\\
&                                 &  Back    		& Back      &  &  &\\
&                                 &  More    		& More      &  &  &\\
&                                 &  ...    		& ...      &  &  &\\
&                                 &  Note    		& Record      &  &  &\\
                                 \cline{1-5}
\multirow{8}{*}{\begin{turn}{90}Mixed Initiative~\cite{radlinski2017theoretical}\end{turn} }  &   & \textbf{Interrupt}      & \textbf{Suggest}      &  &  &\\  
 								\cline{3-5}
&                                & Interrupt      		&       &  &  &\\
&    							&      		& Recommend      &  &  &\\
&                                &      		& Hypothesize     &  &  &\\
&                                &      		&      &  &  &\\
                                \cline{3-5}
&  & \textbf{Interrogate}      & \textbf{Explain}      &  & &\\   
&								& Understand      	&  Report     &  &  &\\
&    							& Explain   		& Reason      &  &  &\\

                      \hline                              
                             
\end{tabular}
\vspace{-4mm}  
\end{table}
\vspace{4mm}

%% file: user-actions.tex
\subsection{User Actions}
During the course of the conversational search process, a user will perform particular actions that will progress the search in different ways. There are two main types of user actions, i.e. those that: (i) change the state of the information need, and those that: (ii) are related to the space of information options/objects (w.r.t the CIN, PINs or AINs). 

\vspace{2mm}
{\bf Reveal Actions}. At various points in the conversation, the user will {\bf disclose} details regarding their information need either voluntarily, or in response to an agent's question, which will then be used to update the CIN, e.g.:
	\begin{center}
	\begin{tcolorbox}[colback=green!5!white,colframe=green!75!black,width=8cm]
    	\scriptsize

		\textbf{User}: I would to arrange a holiday to Italy [Disclose - Volunteer]\\
		\textbf{Agent}: When would to go on holidays?\\
    	\textbf{User}: The 4th of May [Disclose - Inquire].\\
    	...\\
    	\textbf{Agent}: Do you where in Italy you like to go on holidays?\\
    	\textbf{User}: I'm not sure [Disclose - Unsure].\\
        ...\\
    	\textbf{Agent}: What is your budget?\\
    	\textbf{User}: I'd prefer not to say [Disclose - Not].
	\end{tcolorbox}
    \end{center}
    
Of course, there are cases where the user chooses not to disclose information regarding their information need/preferences, either because the user is unsure, or doesn't want to (i.e. {\bf Disclose - Unsure} or {\bf Disclose - Not} ). However, for the search to move forward (or change direction), the user will need to communicate some preferences to the agent.
    
%
%
Since information needs are not fixed requirements will evolve during the course of the search. Consequently, the user may wish to \textbf{revise} and/or \textbf{refine} an existing specified criteria given the CIN. 
Here we assume that an information need is composed of a number of criteria e.g. the holiday destination, the places to visit, the type of holiday, when, duration, which airports to fly to, who is travelling, etc.

%
%

	\begin{center}
	\begin{tcolorbox}[colback=green!5!white,colframe=green!75!black,width=8cm]
    \scriptsize
	\textbf{User}: Actually, we need to go on the 3rd of May in the evening. [Revise (and refine)] \\
    ...\\
    \scriptsize
    \textbf{User}: Do you have any cheap last minute holidays under 300 pounds? [Revise (and Disclose)]
	\end{tcolorbox}
	\end{center}
    In the first example, the user revises their CIN by changing the date, but also adds in another criterion, by asking for an evening departure. Whereas in the second example, the user now decides to further revise their CIN and disclose their budget preference.
    
{\bf Expand}. Rather than constraining the search through a refinement, a user may wish to entertain more possibilities by generalizing or removing the criteria provided.

	\begin{center}
	\begin{tcolorbox}[colback=green!5!white,colframe=green!75!black,width=8cm]
    \scriptsize
	\textbf{User}: Can you also check to see what kinds of holidays are available in Spain? [Expand] \\
     ...\\
     \scriptsize
    \textbf{User}: How about if I increased the budget to 400 pounds? [Expand]
    \end{tcolorbox}
	\end{center}

In the first example, the user now wants to consider both Italy and Spain as possibilities, and then to also consider more expensive holidays. In each case, the search space is now larger (though it may not mean that the number of options available increases).

\vspace{2mm}
{\bf Inquire Actions}. Rather than update or modify the CIN, a common set of related actions involves inquiring about the space of options available given the CIN (or PINs/AINs). As illustrated in the example below,  a user may ask for a list of the different options ({\bf List}), a summary of the different options ({\bf Summarize}), a selection of different options ({\bf Subset}), a comparison between options ({\bf Compare}) or for options that are similar to the current set of options ({\bf Similar}).

	\begin{center}
	\begin{tcolorbox}[colback=green!5!white,colframe=green!75!black,width=8cm]
    \scriptsize
	\textbf{User}: Tell me about all the different things you can do in Tuscany? [Inquire List]\\
    ...\\
    \textbf{User}: Can you give me an overview of the things to do there? [Inquire Summarize]\\
    ...\\
    \textbf{User}: What is the best thing to do in Tuscany?  [Inquire Subset]\\
    ...\\
    \textbf{User}: Which things are suitable for children? [Inquire Subset]\\
    ...\\
    \textbf{User}: What are the main differences between Tuscany and Galicia? [Inquire Compare]\\
    ...\\
    \textbf{User}: What other regions in Europe are like that? [Inquire Similar ]\\
	\end{tcolorbox}
    \end{center}

\vspace{2mm}
{\bf Navigate Actions}. As previously mentioned, conceptually any particular information need has a related set of information objects/options associated with it. The user inquire actions are request to reveal part of this space in some way  i.e. given the set (e.g. all hotels in Montepulciano in Tuscany), in most cases, perform some ordering to form a list (sort by rating/price/etc.), make a selection (pick the best/cheapest/etc.), and then reveal/compare/etc. Consequently, the user may want to navigate within the list in various ways. For example, {\bf Repeat} to revisit the options already revealed, {\bf Back} to go back to previous options, {\bf More} to learn about more options in the list, etc.

During the conversational search process, options will be encountered by the user, that they will want to take {\bf Note} of - that is mark or save in some way, in order to refer back to or consider at a later point in time, and so the user may communicate to the agent something like  ``That hotel could be a possibility.'' or ``Save that hotel for later.''  Given the set of noted options/objects the user will want to revisit this list of options and ask for operations to be performed on this particular set e.g. inquire and navigate.
%
%

\vspace{2mm}
{\bf Interrupt Action}. There will be points in the conversation, for example, when the agent is reciting a long list of items, asking irrelevant questions, etc. when the user will want to interrupt or stop the agent from continuing the conversational such direction. This may be for various reasons - to inspect an item in a list, change the CIN, etc. This interruption request will invariably be coupled with another action/intent, e.g. ``Stop! For the previous hotel, what are the reviews like?''. 

\vspace{2mm}
{\bf Interrogate Actions}. A higher order action of the user is the intent to interrogate the state and understanding of the agent, where the user might like to: (i) know what the agent knows about their information need and the assumptions it has made so far (i.e. {\bf Understand}) , and/or (ii) have the agent explain why particular items are being shown, why particulars suggestions are being made, etc. (i.e. {\bf Explain}).

	\begin{center}
	\begin{tcolorbox}[colback=green!5!white,colframe=green!75!black,width=8cm]
    \scriptsize
	 \textbf{User}: What do you think I am looking for? [Interrogate - Understand]\\
    ...\\
     \scriptsize
    \textbf{User}: Why are you showing me this? [Interrogate - Explain]
    \end{tcolorbox}
    \end{center}

\vspace{2mm}
{\bf Closing Actions}. Finally, the user may choose to end the conversational search process, which could end in various states such as, moving to the task completion stage (``Ok, great, I will book that flight/hotel/car/tour/etc.'' {\bf Complete}. Alternatively, the user may leave with some indication of returning back or without (``Ok, let me think about it - I'll get back to you.'', or ``thanks for your help, bye.''), or by not responding anymore.

%% file: agent-actions.tex
\subsection{Agent Actions }
So far we have explored what actions a user might take when interacting with a conversational search agent. Given these user actions, the agent will need to respond or act accordingly in order to deal with the various requests/responses. 

\vspace{2mm}
{\bf Inquire Actions}. A core action that the agent needs to perform is to {\bf Elicit} the user's information need, by asking the user about various criteria to scope and reduce the possible search space. For very constrained domains this will be based around a template i.e. slot filling, but the bigger challenge is to infer the criteria based on the conversational context and domain. The agent will also need to elicit the user's preferences and constraints.
That is, given a particular criterion how flexible is the criteria? Do they have particular preferences (e.g. ``User: I'd prefer a sunny beach holiday.'') or do they have hard constraints? (e.g. ``User: I definitely do want to go somewhere cold!'').
%
%

So the elicit activity includes: {\bf Elicit Criteria} the different criteria/conditions of the information need, and {\bf Elicit Constraints} the flexibility of the specified criteria/condition. The related task the agent will have to perform is to {\bf Extract} criteria and conditions that have arisen during the conversation.
    
Following on from eliciting criteria and preferences is a need to {\bf Clarify} what the user meant. For example, the agent may want check its understanding due to: miscommunication, i.e. the user was not heard properly (``Did you say you wanted somewhere that is cold?''), or to obtain further specificity (``What do you mean by cold? Less than 20 degrees Celsius?'').


\vspace{2mm}
{\bf Reveal Actions}. In terms of system revealment, the agent will need to disclose information about what it has found given the CIN/PINs/AINs. Either in response to a user list request, or because the agent decides to offer suggestions, the agent will at some point need to{\bf Show} what items it has found by describing the items it has retrieved. 

Given the objects the agent will also need to be able to \textbf{Summarize} or aggregate them to provide the user with an overview of the search space. In the example below we contrast the agent listing the objects vs. summarizing the them available given the user's information need. The agent will need to decide whether it is more appropriate to list objects or summarize (see next subsection).
\begin{center}
	\begin{tcolorbox}[colback=green!5!white,colframe=green!75!black,width=8cm]
    	\scriptsize
		Agent: I've found a number of possible tours around the wine yards. One leaves at 8.30am for 100 pounds, another is at 1.30pm for 75 pounds, and the last one is at 4pm for 139 pounds. [List]
	\end{tcolorbox}
    \end{center}
    
\begin{center}
	\begin{tcolorbox}[colback=green!5!white,colframe=green!75!black,width=8cm]
    	\scriptsize    
		Agent: Tours range from 75 to 139 pounds, and leave in the morning, afternoon and early evening. [ Summarize]
	\end{tcolorbox}
    \end{center}	
    
Other list-based operations that the agent will need to perform include comparing different items in the list, i.e. {\bf Compare}, and to manipulate \textbf{Subsets}, for example, performing operations that will sort, and select some of the objects given a specified criteria (i.e. ``User: What is the cheapest tour?''. The agent will also need to be able to find \textbf{Similar} objects given a particular list objects.

	\begin{center}
	\begin{tcolorbox}[colback=green!5!white,colframe=green!75!black,width=8cm]
    \scriptsize
		\textbf{Agent}: The cheapest tour is for 75 pounds and leaves at 1.30pm, while the evening tour leaves at 4pm and includes a three course dinner, but is more expensive at 139 pounds. [Compare]
    \end{tcolorbox}
    \end{center}

\vspace{2mm}
{\bf Traverse Actions}. Corresponding to the user navigate actions, the agent will need to remember where the user is in the list, and how to update where they are in the search space given their actions (i.e. to \textbf{Repeat}, move \textbf{Back}, provide \textbf{Move} options, etc., and \textbf{Record} when the user shows interest in particular options)

\vspace{2mm}
{\bf Suggest Actions}. At some point in the conversation the agent may either be asked, or may want to make a suggestion regarding the options available. So the agent will need to be able to {\bf Recommend} a particular options/object given the CIN. Alternatively, the agent may want to {\bf Hypothesize} about the CIN and generate what-if information needs i.e. (AINs) such as:
what-if they could go to a different country/region/place, what if they could go on different dates, etc. These Alternative Information Needs (AINs) may lead to a different space of options which might (or might not) be of relevance to the user. The subtle distinction here is whether the suggestion is based on what is currently being discussed (i.e. the CIN) , or whether it is based on some variation (i.e. an AIN).

\vspace{2mm}
{\bf Explain Actions}.  The agent will need to be able to \textbf{report} its understanding of the CIN to respond to a user's interrogation or to provide context to the user. Further, it will also need to be able to justify and \textbf{reason} why it took a particular course of action, made a particular suggestions, etc.

	\begin{center}
	\begin{tcolorbox}[colback=green!5!white,colframe=green!75!black,width=8cm]
    \scriptsize
	\textbf{Agent}: Ok, so you are interested in a sunny holiday in a place where you can explore different interesting cultural sites. Is that correct? [Report]\\
	\textbf{User}: Yes.
   \tcblower
   \scriptsize
   \textbf{User}: Why did you recommend going to Tuscany?\\
   \textbf{Agent}: Tuscany is a beautiful region of Italy known for hot days and warm nights, and has a variety of interesting sites to visit with cultural significance. [Reason]
       \end{tcolorbox}
    \end{center}
            
{\bf Error and Finalization Actions}.           
When an agent gets confused or misunderstands what the user wants it will need to \textbf{recover} smoothly. This may be by updating the information need to reflect what the user actually wanted, or to go back in the conversation where they both share a common understanding of the search space. The agent will also need to {\bf handle} requests/actions that are outside of the agents scope or capacity. For example, when a user about medical conditions when it is designed to help plan holidays, etc.

{\bf End}. At some point, the conversation will draw to a close - and it may or may not be resumed. For example, the user might have decided on the various aspects of their holiday (and thus will transition to the booking agent). On the other hand, the user might revisit or continue exploring the space of options, before and even during the holiday itself. The agent will therefore need to track and persist the conversation, and decide how to respond depending on the context.

%% file: decision-points.tex
\subsection{Agent Decisions and Tasks}
So far we have enumerated different actions that users and agents may perform during the conversational search process - and shown in Table~\ref{tbl_actions} how they relate to one another. In this subsection, we aim to discuss in more detail the different decisions and tasks that the agent needs to perform to facilitate the conversation.

{\bf Agent Dialog Policy}. The overarching decision that the agent needs to perform is to decide on how to respond/initiate given the the preceding conversation and the user's request/response. The agent must decide on what action or actions to take in order to provide a useful and meaningful response that drives the conversation forward (w.r.t the user's goals). Assuming a turn-based dialogue model, where the user initiates the dialogue (e.g., ``I'm looking to go on holidays to Italy.''), what action(s) should the agent take --- at the high level --- the agent can: \textit{inquire}, \textit{reveal}, \textit{traverse}, \textit{suggest} or \textit{explain}. The Agent's dialog policy will define how the agent will act and behave based on the context of the conversation. However, some actions are going to be better than others --- better in the sense that they advance the user towards their goal in some way, and do so in an efficient manner. Essentially the agent needs to decide what action(s) it should take next. For example, the agent could: (a) inquire-elicit, (b) reveal-list, (c) reveal-summarize, (d) explain-report, etc. Crafting and/or inferring the dialog policy is an open challenge and core task of a conversational search agent. 
Below, we describe a number of the lower level tasks that will be part of such a policy. 





\vspace{2mm}
{\bf Intent Identification and Extraction Task}. A key task performed during each turn is to extract out the intent(s) of the user given their utterance. 
For example, given the user's initial statement above, the agent needs to identify the intent of the user, in this case they are revealing and disclosing their information need i.e. \textit{Reveal-Disclose}. Given this intention, the agent then needs to \textit{extract} out the disclosed criteria and represent this within the current information need, i.e.. the CIN models that the user wants to go on holiday to Italy.  Being able to infer and extract the intention will be a core task that any conversational agent will need to be able to perform accurately.


\vspace{2mm}
{\bf Inquire-Elicit Question Selection Task}. For a given CIN, the search space may be quite large or have many different aspects/facets which represent different ways to explore and narrow down the search space. For example, given the current example, there are hundreds of thousands of possible holiday options. Thus, an agent will need to decide what question to ask the user in order to narrow down the search space to help refine the CIN. Of course, there is a range of questions that the agent could ask. For example, where in Italy, when they would like to travel, what they want to do, how much they are willing to spend, who will be going, etc. Given the agent's objectives, the agent will most likely want to ask a question that reduces the search space (such that answer from the user will result in the number of options associated with the updated information need less than the number of options associated with the previous information need).
However, there may be cases when/if the information need is over specified, and so the agent will need to ask question which requires the user to expand their information need. A key challenge here is to select a question that efficiently decomposes the search space, but is also contextually relevant i.e. makes sense to the user, and maximizes the user's understanding of the search space. 

\vspace{2mm}
{\bf Inquire-Clarify Requirement Clarification Task}. A related task that the agent will need to perform is to decide whether it should seek a clarification, given the previous utterance (and what the agent has inferred and extracted from it). This may arise in a number of situations: the agent doesn't understand what the user said, or the utterance doesn't make sense in some way, or the information provided is under specified. For example, let's assume the agent asks the user when they would like to travel, and the user responds, ``On the 4th''.  Does the user mean ``May the 4th'', ``June the 4th'', etc. it is ambiguous. The agent must decide whether to: (i) leave it unspecified, (ii) impute the missing details (assume they meant ``May the 4th, 2018'', or (iii) ask a clarifying question  e.g.``Do you mean the 4th of May, 2018 ?''.  Depending on how much the agent understands of the domain, and the given context, will determine how much the agent can assume or impute given the underspecified criteria. If the agent doesn't make any assumptions, and keeps asking clarifying questions, then the conversation may get unnecessarily bogged down in details. On the other hand, if too many assumptions are made, the agent's representation of the information need may significantly differ from the user's representation. 
This will mean that there may be a trade-off between the efficiency of the conversation and the accuracy of the information need as the agent has to decide between how important it is to clarify and how risky it is to infer or impute the underspecified or missing details.

\vspace{2mm}
{\bf Elicit-Reveal Decision Task}. At some point during the conversation, the agent will need to decide whether it should: (i) continue eliciting requirements from the user, or (ii) given the CIN reveal options to the user (in some manner). For example, let's assume that the CIN is that the user wants to visit Nice in France, and wants to know about the different museums and galleries in the area. To go through and list all options, for most use cases, is not going to be a very efficient, and quite possibly quite cognitively taxing on the user. Conversely, if the CIN is so overly specified that only a few or no options are available, then eliciting further requirements is also inefficient and pointless. At some point, the time it takes to elicit will be similar to the time it takes to reveal. For example, it may be more efficient and useful if agent reveals: ``There are variety of museums and galleries depending on what your interested in, specific artists, local and living artists, modern art, popular or period exhibitions'' rather asking the user about the different aspects, one at a time. Finding the balance between eliciting and revealing is a key decision that the agent will have to make, which will again influence how efficient and effective the agent is in helping the user explore the space of available options.


{\bf Reveal Task}. Once the agent decides to reveal options, it then needs to decide how to reveal the objects available to the user, i.e. should the agent: list, summarize, compare, etc. Again how the options are revealed will impact the direction and efficiency of the conversation. For example, providing a long listing of the options will take longer and be more cognitively taxing than providing a summary of the set of options, so that the user can drill down. A further key challenge here is to provide a compact, descriptive and useful summary, listing or comparison of the objects that focuses on the most salient features that are relevant to the task at hand i.e. helping the user understand the possible space of options available so that they can further refine/expand their information need accordingly.

{\bf Suggest Task}. During the conversation, the agent may decide to make suggestions based on the current or past information needs. As such the agent needs to generate hypothetical i.e. alternate information needs. For example, the user may have narrowed down the search space to a handful of museums to visit. The agent might then decide to provide alternatives, such as other related museums, ones that are close to the others, or museums that are similar but rated more highly. This presents yet another conversational trade-off --- suggestions help provide the user with a better understanding of the space of options (i.e. maximizes exploration), but at the expense of increasing the conversational effort. If the agent keeps suggesting alternative options, the conversation will be protracted and drawn out, frustrating the user especially if non-relevant options are been recommended. However, the user may be rather dissatisfied if they later learn that a better option was available to them that the agent had not suggested.

{\bf Report Task}. During the conversation, the user may explicitly ask the agent to report their understanding of the CIN, however, the agent may also decide to voluntarily disclose what it understands. One one hand, reporting the state will provide conversational awareness ensuring that the user and agent are on talking about the same thing, but it will elongate the conversation and increase the conversational effort. On the other hand, if the agent does not report its understanding, it risks straying from what the user actually wants. So two challenge that arises here is how to efficiently report the agent's understanding, and when to report its understanding to the user.

Above we described a number of tasks that the agent is faced with in processing and dealing with the user's requests/responses given the conversation. This is only a subset of core tasks that agents need to be able to perform. By breaking down the conversational process into these tasks, we can focus more specifically on the different decisions that agent needs to make and consider how these actions can be evaluated separately. 

%% file: closing.tex
\section{Discussion}\label{sec_discussion}
In this work, we enumerate key aspects of the information action space for users and agents during the conversational search process. We created a simple conceptual framework for conversational search by providing a set of possible actions that agents need to perform and the key decisions that they have to make during the conversational search process. 

While the conceptual framework is an incomplete specification, it provides a starting point for the development of complete interaction model, that is based on previous work. By outlining these different actions together we begin to obtain a clearer picture of the nature of the conversational search process -- and in doing so we have identified various trade-offs between different actions. For example, the contrasting objectives to minimize effort while also maximizing exploration of relevant objects. 

In this paper, we have not discussed nor specified how to implement the actions or decisions that the agent needs to perform. This is very much an open problem. However, the conceptual representation helps to draw attention to these key decisions and actions, so that implementations of the interaction model will allow aspects of the conceptual framework to be tested empirically, and identify which actions/intents are critical and which ones may be discarded or refined. 

We hope that the proposed framework will lead to a more principled approach to the  development  of conversational search agent because the community will be able to focus on th key actions and intents as well as the process (decision points) and design evaluation tasks to examine them in detail. As the sophistication of agents develop, so too will the space of actions capable of being performed. To support each action and intent is a challenge in its own right - for example - how should the agent reveal options/objects, summarize a set of options, etc., and when should the agent elicit, reveal, explain, hypothesize, etc. So while, we make no claim that this is a definitive conceptual framework. Its goal is to serve as a starting point for deeper discussion, such as:
\begin{itemize}
\item What other actions, intents and decisions should be considered?  
\item How can we best represent the interaction between the agent and user in such a framework? 
\item How can we more formally represent and model the interaction? 
\item And, how can we model the state of the information needs (the current information need, the possible information needs, etc.) and their influence on the search space? 
\end{itemize}

Going beyond the workshop we will look to how we can empirically validate the framework using conversational search logs with humans (e.g.~\cite{mcduff2017multimodal}) and how we can develop a software based framework to support our conceptualization.